\documentclass[fleqn,usenatbib]{mnras}

\usepackage{newtxtext,newtxmath}

\usepackage{tensor}
\usepackage{multirow,array}
\usepackage{booktabs} 

\usepackage[T1]{fontenc}

\DeclareRobustCommand{\VAN}[3]{#2}
\let\VANthebibliography\thebibliography
\def\thebibliography{\DeclareRobustCommand{\VAN}[3]{##3}\VANthebibliography}

\usepackage{graphicx}	
\usepackage{amsmath}	

\title[Constraining meVSL by CC]{Constraining minimally extended varying speed of light by cosmological chronometers}

\author[Seokcheon Lee]{
Seokcheon Lee,$^{1}$\thanks{E-mail: skylee@skku.edu}
\\
$^{1}$ Department of Physics, Institute of Basic Science, Sungkyunkwan University, Suwon 16419, Korea
}

\date{Accepted XXX. Received YYY; in original form ZZZ}

\pubyear{2023}

\begin{document}
\label{firstpage}
\pagerange{\pageref{firstpage}--\pageref{lastpage}}
\maketitle

\begin{abstract}
At least one dimensionless physical constant (\textit{i.e.}, a physically observable) must change for the cosmic time to make the varying speed of light (VSL) models phenomenologically feasible. Various physical constants and quantities also should be functions of cosmic time to satisfy all known local laws of physics, including special relativity, thermodynamics, and electromagnetism. Adiabaticity is another necessary condition to keep the homogeneity and isotropy of three-dimensional space. To be a self-consistent theory, one should consider cosmic evolutions of physical constants and quantities when one derives Einstein’s field equations and their solutions. All these conditions are well satisfied in the so-called minimally extended varying speed of light (meVSL) model. Unlike other VSL models, we show that the redshift-drift formula of the meVSL model is the same as a standard model. Therefore, we cannot use this as an experimental tool to verify the meVSL. Instead, one can still use the cosmological chronometers (CC) as a model-independent test of the meVSL. The current CC data cannot distinguish meVSL from the standard model (SM) when we adopt the best-fit values (or Gaussian prior) of $H_0$ and $\Omega_{m0}$ from the Planck mission. However,  the CC data prefer the meVSL when we choose Pantheon22 data. 
\end{abstract}




\section{Introduction}

Physically it makes sense to consider the variability of dimensionless constants only. Nonetheless, any physical theory requires dimensional parameters. Thus, experimental results are only sensitive to dimensionless combinations of dimensional constants (quantities). Also, any theory with the variation of any dimensional constant should include variations of dimensionless quantities \citet{2000ApJ...532L..87B}.  Any indication of the cosmic variation of physical constants would provide important implications for cosmology.

In the meVSL model, the fine structure constant ($\alpha \equiv e^2/(4 \pi \epsilon \hbar c)$, one of the dimensionless physical constants ) also varies as a function of cosmic time when the speed of light varies for cosmic time.  Physical constants and quantities of this model also evolve to satisfy all known local physics laws, including special relativity, thermodynamics, and electromagnetism as shown in Tab.~\ref{tab:table-1} \citet{2021JCAP...08..054L}. Recently, it has shown there is no theoretical contradiction of the cosmic time evolution of the speed of light based on the Robertson-Walker metric. Therefore, it becomes important to prove its validity from cosmological observation \citet{2023FoPh...53...40L}.

\begin{table*}
	\centering
\begin{tabular}{|c||c|c|c|}
	\hline
	local physics laws & Special Relativity & Electromagnetism & Thermodynamics \\
	\hline \hline
	quantities & $\tilde{m} = \tilde{m}_0 a^{-b/2}$ & $\tilde{e} = \tilde{e}_0 a^{-b/4}\,, \tilde{\lambda} = \tilde{\lambda}_0 a \,, \tilde{\nu} = \tilde{\nu}_0 a^{-1+b/4}$ & $\tilde{T} = \tilde{T}_0 a^{-1}$ \\
	\hline
	constants & $\tilde{c} = \tilde{c}_0 a^{b/4} \,, \tilde{G} = \tilde{G}_0 a^{b}$ & $\tilde{\epsilon} = \tilde{\epsilon}_0 a^{-b/4} \,, \tilde{\mu} = \tilde{\mu}_0 a^{-b/4} \,, \tilde{c} = \tilde{c}_0 a^{b/4}$ & $\tilde{k}_{\textrm{B} 0} \,, \tilde{\hbar} = \tilde{\hbar}_0 a^{-b/4} \,, \tilde{c} = \tilde{c}_0 a^{b/4}$ \\
	\hline
	energies & $\tilde{m} \tilde{c}^2 = \tilde{m}_0 \tilde{c}_0^2$ & $\tilde{h} \tilde{\nu} = \tilde{h}_0 \tilde{\nu}_0 a^{-1}$ & $\tilde{k}_{\textrm{B}} \tilde{T} = \tilde{k}_{\textrm{B} 0} \tilde{T}_0 a^{-1}$ \\
	\hline
\end{tabular}
\caption{Summary for cosmological evolutions of physical constants and quantities of the meVSL model. These relations satisfy all known local physics laws, including special relativity, thermodynamics, and electromagnetism.}
\label{tab:table-1}
\end{table*}

Even though Sandage proposed the possibility of using the time variation of the redshift of a source to probe cosmological models \citet{1962ApJ...136..319S}, its predicted signal was deemed too small to be measured. However, its observation becomes feasible due to the foreseen development of large observatories, such as the European Extremely Large Telescope (E-ELT),  the Square Kilometer Array (SKA) \cite{1998ApJ...499L.111L,2008MNRAS.386.1192L,2015aska.confE..27K}, and  Five-hundred-meter Aperture Spherical radio Telescope (FAST) \citet{2022PDU....3701088L}.  Depending on the underlying cosmological model, the redshift of any given object shows a specific variation in time (\textit{i.e.}, redshift drift (RD)). It is a direct probe of the dynamics of the expansion free from the calibration of standard candles or a standard ruler. It is also independent of the geometry of the Universe. Thus, it might be used as a valuable tool to support the accelerated expansion independently or to investigate the dynamical behavior of the expansion expected in general relativity (GR) compared to alternative scenarios.  This proposal was considered in some VSL models \citet{2014PhLB..728...15B,2016PhRvD..93f3521S,2017PhRvD..95h4035S}.  It is an expanding research area because of its usefulness \citet{2008PhRvD..77b1301U,2010PhLB..692..219J,2012PhR...521...95Q,2011JCAP...07..008G,2011PhRvD..84j4003M,2013ApJ...769..133N,2012PhRvD..86l3001M,2012ApJ...761L..26D,2015mgm..conf.1590M,2013JCAP...08..030A,2014MmSAI..85...13M,2013arXiv1311.6858Z,2015APh....62..195K,2015PhRvD..91f3006L,2015FrPhy..10j9501G,2016PhRvD..94d3001M,2016MNRAS.463L..61M,2017PhRvD..95j1301P,2018JCAP...04..002J,2018EPJC...78...11L,2018PhRvD..98b3517M,2019EPJC...79..267T,2019BAAS...51c..53E,2020JCAP...01..054J,2019arXiv190704495B,2019MNRAS.488.3607A,2020EPJC...80..304L,2019BAAS...51g.137E,2020MNRAS.492.2044C,2020JCAP...04..043L,2020NewA...8101425B,2021PhRvD.103b3537H,2021JCAP...12..042Q,2021PhRvD.103h1302H,2021PhRvL.126w1101K,2021arXiv210704868C,2021PhRvD.104l3527H,2021MNRAS.508L..53E,2021MNRAS.508.3620W,2022EJPh...43c5601M,2022LRR....25....6M,2022arXiv220305924C,2022MNRAS.513.5198C,2022PhRvD.106d3501K,2022MNRAS.514.5493D,2023NewA...9901940S,2022ApJ...940...16W,2022arXiv221013946L,2023MNRAS.518.2853R,2023MNRAS.519.2769H}.  

Cosmic chronometers (CC) are other direct measurements of the Hubble parameter using two passively evolving (elliptical) galaxies assumed to have formed at the same cosmic epoch but observed at different redshifts \citet{2002ApJ...573...37J,2011JCAP...03..045M}. This approach also does not rely on integrated quantities, such as the luminosity distance,  and is free from the pre-assumption of any particular model 
\citet{2017ApJ...835..270W,2017MNRAS.467.3239R,2020ApJ...888...99W,2020ApJ...898...82M,2021ApJ...908...84V,2021MNRAS.506L...1D,2022ApJ...927..164B,2022ApJ...928L...4B,2022arXiv220902426B,2022EPJP..137.1341J,2022arXiv221203821A,2022arXiv221205731K,2022arXiv221214291L}. Thus, it is good to use for investigating VSL models \citet{2022JCAP...07..029R}. 

This paper is organized as follows. In section \ref{sec:meVSL}, we review the basics of the meVSL model and its properties. In section \ref{sec:Dmeadotz}, we derive formulae for the RD and CC of the meVSL model. We also analyze it for the meVSL model.  We give our conclusion in section \ref{sec:Conc}.

\section{Review of meVSL}
\label{sec:meVSL}

The application of the Robertson-Walker (RW) metric to GR led to the success of modern cosmology. The so-called SM is considered a flat $\Lambda$CDM model. However, despite the success of the SM, it is also true that there is still tension between cosmological parameters in some observations. Therefore, it is crucial to reconsider and test the alternative of the SM or its foundation. As an alternative, we propose the so-called meVSL model in which physical constants vary over cosmic time, which still satisfies local electromagnetism and thermodynamics \citet{2021JCAP...08..054L}. We briefly review this model in this section.

\subsection{Hilbert Einstein action}\label{subsec:HEaction}

We obtain Einstein's field equation (EFE) through the principle of least action from the Einstein-Hilbert (EH) action. If one only allows the speed of light to vary as a function of cosmic time, it causes the problem in recovering the EFE due to the Palatini identity term acting on the varying speed of light. Thus, one also should allow the gravitational constant to vary such that the combination of them ({\it i.e.}, $\tilde{\kappa} \equiv 8 \pi \tilde{G}/\tilde{c}^4$) in the EH action does not depend on the cosmic time. From this point, one writes the EH action of meVSL as
\begin{align}
S &\equiv \int \Biggl[ \frac{1}{2 \tilde{\kappa}} \left( R - 2 \Lambda \right) + \mathcal L_{m} \Biggr] \sqrt{-g} dt d^3x \label{SHmpApp} \,,
\end{align}
where $\tilde{\kappa}$ is the so-called Einstein gravitational constant. The variation of the action with respect to the inverse metric should be zero 
\begin{align}
\delta S &= \int \frac{\sqrt{-g}}{2 \tilde{\kappa}} \left[ R_{\mu\nu} - \frac{1}{2} g_{\mu\nu} \left( R - 2 \Lambda \right) - \tilde{\kappa} T_{\mu\nu} \right] \delta g^{\mu\nu} dtd^3 x \nonumber \\ &+ \int \frac{\sqrt{-g}}{2 \tilde{\kappa}} \left[ \nabla_{\mu} \nabla_{\nu} - g_{\mu\nu} \Box \right] \delta g^{\mu\nu} dtd^3 x = 0 \label{deltaSHmp2App} \,.
\end{align}
In order not to spoil the EFE, the second term ({\it i.e.}, Palatini identity term) in the above Eq.~\eqref{deltaSHmp2App} should vanish. It means that $\tilde{\kappa}$ should be constant even though both $\tilde{c}$ and $\tilde{G}$ cosmologically evolve. 
\begin{align}
&\tilde{\kappa} = \textrm{const} \quad \Rightarrow \quad \tilde{c}^4 \propto \tilde{G} \nonumber \,, \\
& \tilde{c} = \tilde{c}_0 a^{b/4} \quad , \quad \tilde{G} = \tilde{G}_{0} a^{b} \label{tkappaconstmpApp} \,, 
\end{align}
where we put $a_0 = 1$, and $\tilde{c}_0$ and $\tilde{G}_0$ denote present values of the speed of light and the gravitational constant, respectively. From the above two equations \eqref{deltaSHmp2App} and \eqref{tkappaconstmpApp}, one can obtain the EFE including the cosmological constant
\begin{align}
&R_{\mu\nu} - \frac{1}{2} g_{\mu\nu} R + \Lambda g_{\mu\nu} \equiv G_{\mu\nu} + \Lambda g_{\mu\nu}  = \frac{8 \pi \tilde{G}}{\tilde{c}^4} T_{\mu\nu} \label{tEFEmpApp} \;, 
\end{align}
where $G_{\mu\nu}$ is the Einstein tensor. The above EFE has the same form as that of SM. 

\subsection{RW metric}\label{subsec:FLRWmppp}
We now derive the Friedmann equations of the meVSL model based on the RW metric. The RW metric is a spatially homogeneous and isotropic spacetime given by
\begin{align}
ds^2 &= g_{\mu\nu} dx^{\mu} dx^{\nu} = -\tilde{c}^2 dt^2 + a^2 \gamma_{ij} dx^i dx^j \nonumber \\ &= -\tilde{c}^2 dt^2 + a^2 \left( \frac{dr^2}{1-k r^2} + r^2 d \Omega^2 \right)\label{ds2App} \,.
\end{align} 
The Christoffel symbols for the RW metric in Eq.~\eqref{ds2App} are given by
\begin{align}
&\Gamma^{0}_{ij} = \frac{a\dot{a}}{\tilde{c}} \gamma_{ij} \quad , \quad \Gamma^{i}_{0j} = \frac{1}{\tilde{c}}  \frac{\dot{a}}{a} \delta^i_j \quad , \quad \Gamma^{i}_{jk} = ^{s}\Gamma^{i}_{jk}  \label{GammacompApp} \,,
\end{align}
where $^{s}\Gamma^{i}_{jk}$ denotes the Christoffel symbols for the spatial metric $\gamma_{ij}$. As shown in the above Eq.~\eqref{GammacompApp}, the Christoffel symbols of meVSL are the same forms as those of SM. However, $\tilde{c}$ varies as a function of the scale factor.

The curvature of the Riemann manifold is expressed by the Riemann curvature tensors that are given by
\begin{align}
&\tensor{R}{^0_i_0_j} = \frac{\gamma_{ij}}{\tilde{c}^2} \left( a \ddot{a} - \dot{a}^2 \frac{d \ln \tilde{c}}{d \ln a} \right) \quad , \quad \tensor{R}{^i_0_0_j} = \frac{\delta^{i}_{j}}{\tilde{c}^2} \left( \frac{\ddot{a}}{a} - \frac{\dot{a}^2}{a^2} \frac{d \ln \tilde{c}}{ d \ln a}  \right) \,, \label{R0i0jApp} \\
&\tensor{R}{^i_j_k_m} = \frac{\dot{a}^2}{\tilde{c}^2} \left( \delta^{i}_{k} \gamma_{jm} - \delta^i_m \gamma_{jk} \right) + \tensor[^s]{R}{^i_j_k_m} \quad , \nonumber \\
& \tensor[^s]{R}{^i_j_k_m} = k \left( \delta^i_k \gamma_{jm} - \delta^i_m \gamma_{jk} \right) \,. \label{RijkmApp} 
\end{align}
Even though the Christoffel symbols of the RW metric of the meVSL model are the same form as those of SM, the Riemann curvature tensors of meVSL are different from those of SM. It is because the Riemann curvature tensors are obtained from the derivatives of the Christoffel symbols including the time-varying speed of light with respect to the cosmic time $t$ ({\it i.e.}, the scale factor $a$). Thus, one obtains the correction term ($H^2 d \ln \tilde{c}/ d \ln a$) in both $\tensor{R}{^0_i_0_j}$ and $\tensor{R}{^i_0_0_j}$. The Ricci curvature tensors measuring how a shape is deformed as one moves along geodesics in the space are obtained from the contraction of the Riemann curvature tensors given in Eqs.~\eqref{R0i0jApp} and \eqref{RijkmApp}
\begin{align}
R_{00} &= -\frac{3}{\tilde{c}^2} \left( \frac{\ddot{a}}{a} - \frac{\dot{a}^2}{a^2} \frac{d \ln \tilde{c}}{ d \ln a}  \right) \,, \nonumber \\
R_{ij} &= \frac{\gamma_{ij}}{\tilde{c}^2} a^2 \left( 2 \frac{\dot{a}^2}{a^2} + \frac{\ddot{a}}{a} + 2 k \frac{\tilde{c}^2}{a^2} - \frac{\dot{a}^2}{a^2} \frac{d \ln \tilde{c}}{ d \ln a}  \right) \label{RijApp} \,. 
\end{align}
Again, there is a correction term in both $R_{00}$ and $R_{ij}$ due to the time-variation of the speed of light.  Finally, one can also obtain the Ricci scalar by taking the trace of the Ricci tensors
\begin{align}
R &= \frac{6}{\tilde{c}^2} \left( \frac{\ddot{a}}{a} + \frac{\dot{a}^2}{a^2} + k \frac{\tilde{c}^2}{a^2} - \frac{\dot{a}^2}{a^2} \frac{d \ln \tilde{c}}{ d \ln a}  \right) \label{RmpApp} \,,
\end{align} 
where the time-varying speed of the light effect is shown in the last term. 

\subsection{Stress Energy Tensor}\label{subsec:Tmunu}

To solve the EFE given in Eq.~\eqref{tEFEmpApp}, one needs the stress-energy tensor in addition to the geometric terms in Eqs.~\eqref{RijApp} and \eqref{RmpApp}. The symmetric stress-energy tensor of a perfect fluid in thermodynamic equilibrium, which acts as the source of the spacetime curvature, is given by
\begin{align}
T_{\mu\nu} = \left( \rho + \frac{P}{\tilde{c}^2} \right) U_{\mu} U_{\nu} + P g_{\mu\nu} \label{TmunumpApp} \;,
\end{align}
where $\rho$ is the mass density and $P$ is the hydrostatic pressure. The covariant derivatives of both the Einstein tensor $G_{\mu\nu}$ and the metric $g_{\mu\nu}$ are zero (\textit{i.e.}, the Bianchi identity). From the Bianchi identity and the constancy of the Einstein gravitational constant $\tilde{\kappa}$, one can obtain the local conservation of energy and momentum as
\begin{align}
&\tensor{T}{^\mu_\nu_;_\mu} = 0 \quad \Rightarrow \quad \frac{\partial \rho_i}{\partial t} + 3 H \left( \rho_i + \frac{P_i}{\tilde{c}^2} \right) + 2 \rho_i H \frac{d \ln \tilde{c}}{d \ln a} = 0 \label{BI1mpApp} \;.
\end{align}
One can solve for this equation~\eqref{BI1mpApp} to obtain
\begin{align}
&\rho_i \tilde{c}^{2} = \rho_{i0} \tilde{c}_0^2 a^{-3 (1 + \omega_i)} \quad , \quad
\rho_{i} = \rho_{i0} a^{-3(1 + \omega_i) -\frac{b}{2}} \label{rhompApp} \;,
\end{align}
where we use the equation of state $\omega_i = P_i / (\rho_i \tilde{c}^2)$ for the $i$-component. The mass density of $i$-component redshifts slower (faster) than that of SM for the negative (positive) value of $b$. Thus, one might interpret Eq.~\eqref{rhompApp} as the rest mass evolves cosmologically $a^{-b/2}$. 

\subsection{FLRW universe}\label{subsec:FLRWUniv}

By inserting Eqs.~\eqref{RijApp}, ~\eqref{RmpApp}, ~\eqref{TmunumpApp}, and ~\eqref{rhompApp} into Eq.~\eqref{tEFEmpApp}, one can obtain the components of EFE
\begin{align}
&\frac{\dot{a}^2}{a^2} + k \frac{\tilde{c}^2}{a^2}  -\frac{ \Lambda \tilde{c}^2}{3} = \frac{8 \pi \tilde{G}}{3} \sum_i \rho_i \label{tG00mpApp} \,, \\ 
&\frac{\dot{a}^2}{a^2} + 2 \frac{\ddot{a}}{a} +  k \frac{\tilde{c}^2}{a^2} - \Lambda \tilde{c}^2 - 2 \frac{\dot{a}^2}{a^2} \frac{d \ln \tilde{c}}{d \ln a} = -\frac{8 \pi \tilde{G}}{\tilde{c}^2} \sum_{i} P_i  \label{tG11mpApp} \;. 
\end{align} 
One obtains the expansion acceleration from Eqs.~\eqref{tG00mpApp} and \eqref{tG11mpApp}
\begin{align}
&\frac{\ddot{a}}{a} = -\frac{4\pi \tilde{G}}{3} \sum_i \left( 1 + 3 \omega_i \right) \rho_i  + \frac{\Lambda \tilde{c}^2}{3} + \frac{\dot{a}^2}{a^2} \frac{d \ln \tilde{c}}{d \ln a}\label{t3G11mG00mpApp} \,. 
\end{align} 
One can rewrite the Hubble parameter $H$ and the acceleration $\ddot{a}/a$ by using Eqs.~\eqref{tkappaconstmpApp} and \eqref{rhompApp} as
\begin{align}
H^2 &= \left[ \frac{8 \pi \tilde{G}_0}{3} \sum_{i} \rho_{0i} a^{-3(1+\omega_i)} + \frac{ \Lambda \tilde{c}_0^2}{3} - k \frac{\tilde{c}_0^2}{a^2} \right] a^{\frac{b}{2}} \nonumber \\ &\equiv H^{(\textrm{SM})2} a^{\frac{b}{2}} \equiv H_0^2 E^{2} a^{\frac{b}{2}}  \label{H2meApp} \;, \\
\frac{\ddot{a}}{a} &= \left[ -\frac{4\pi \tilde{G}_0}{3} \sum_i \left( 1 + 3 \omega_i \right) \rho_{0i} a^{-3(1+\omega_i)} + \frac{\Lambda \tilde{c}_0^2}{3} \right] a^{\frac{b}{2}}  + H^2 \frac{d \ln \tilde{c}}{d \ln a} \nonumber \\
&= \left[ \left( \frac{\ddot{a}}{a} \right)^{(\textrm{SM})} + \frac{b}{4} H^{(\textrm{SM})2} \right] a^{\frac{b}{2}} \,,
\end{align}
where $E(z) \equiv H^{(\textrm{SM})}(z)/H_0 = \sqrt{\Omega_{\textrm{m} 0} (1+z)^{3} + (1 - \Omega_{\textrm{m} 0})}$ for the flat $\Lambda$CDM model. $H^{(\textrm{SM})}$ and $ ( \ddot{a}/a )^{(\textrm{SM})}$ denote the Hubble parameter and the expansion acceleration for the constant speed of light, respectively. These equations are background evolutions of the FLRW universe of meVSL model. The expansion speed of the Universe in meVSL, $H$ has the extra factor $(1+z)^{-b/4}$ compared to that of SM, $H^{(\textrm{SM})}$. Thus, the present values of the Hubble parameter of SM and meVSL are the same. However, the value of the Hubble parameter of meVSL in the past is $(1+z)^{-b/4}$-factor larger (smaller) than that of SM if $b$ is negative (positive). This simple fact might be used to solve the Hubble tension.


\subsection{Cosmological redshift}\label{subsec:redshiftz}

Now we consider a light reaching us, at $r = 0$, has been emitted from a galaxy at $r=r_1$. Also, we consider successive crests of light, emitted at times $t_1$ and $t_1 + \Delta t_1$ and received at times $t_0$ and $t_0 + \Delta t_0$. Since $ds^2 = 0$ and the light is traveling radially one has for the first and the second crest of light
 \begin{align}
 \int_{r_1}^{0} \frac{dr}{\sqrt{1 - kr^2}} &=  \int_{t_1}^{t_0} \frac{\tilde{c} dt}{a} = \int_{t_1+\Delta t_1}^{t_0+\Delta t_0} \frac{\tilde{c} dt}{a} \label{t1t2}
\,. \end{align} 
One can rewrite the above equation as
 \begin{align}
\int_{t_0}^{t_0+\Delta t_0} \frac{\tilde{c} dt}{a} &= \int_{t_1}^{t_1+\Delta t_1} \frac{\tilde{c} dt}{a}   \quad \Longrightarrow \quad   \frac{\tilde{c}_0 \Delta t_0}{a_0} =  \frac{\tilde{c}_1 \Delta t_1}{a_1} \label{Deltaa} \,,
 \end{align} 
where we use $\Delta t_1$ and $\Delta t_0$ are very small ({\it i.e.}, $\Delta t \ll t$).  Eq.~\eqref{Deltaa} provides \citet{2000ApJ...532L..87B,2021JCAP...08..054L}
 \begin{align}
 \frac{\Delta t_1}{a(t_1)^{1-b/4}} =  \frac{\Delta t_0}{a(t_0)^{1-b/4}} \quad \Rightarrow \quad \frac{1}{a(t_1) \tilde{\nu}(t_1)} = \frac{1}{a(t_2) \tilde{\nu}(t_2)}  \label{redshiftmp} \,.
 \end{align}
Cosmological redshift is the relative difference between observed and emitted wavelengths of an object. The cosmological redshift $z$ represents this change.  If $\tilde{\lambda}$ represents wavelength and $\tilde{\nu}$ does frequency,  one can define $z$ from the difference between the emitted and observed wavelengths. Both the emitted and observed wavelengths are given by
 \begin{align}
 \tilde{\lambda}_{\textrm{e}} \equiv \frac{\tilde{c}(t_1)}{\tilde{\nu}(t_1)} = \frac{\tilde{c}_0}{\tilde{\nu}_0} \frac{a_{1}}{a_0}  \quad , \quad \tilde{\lambda}_{\textrm{o}} \equiv \frac{\tilde{c}_0}{\tilde{\nu}_0} \label{lambdaeo} \,.
 \end{align}
 If we use Eqs.~\eqref{redshiftmp} and ~\eqref{lambdaeo}, then we obtain
 \begin{align}
1 + z \equiv \frac{\tilde{\lambda}_{\textrm{o}}}{\tilde{\lambda}_{\textrm{e}}} = \frac{a_0}{a_1}  \label{redshiftmp2} \,.
 \end{align}
 Thus, the redshift, $z$ in meVSL is the same as that of SM. This fact is important because cosmological observations are expressed by using the redshift, not by cosmic time. If the $z$ of any VSL model is different from that of SM, then one needs to reinterpret the observational data by using a new redshift obtained from that model. In meVSL, as $\tilde{c}$ changes as a function of time, so does the frequency. One can use $\tilde{c} = \tilde{\nu} \tilde{\lambda}$ where $\tilde{\nu} = \tilde{\nu}_0 a^{-1+b/4}$ by using Eq.~\eqref{redshiftmp}. 
 
\subsection{Hubble radius}\label{subsec:coH}

One of the main motivations of previous VSL models is providing the model alternative to cosmic inflation by shrinking the so-called comoving Hubble radius in time ({\it i.e.}, $d (c/aH)/dt < 0$). However, one obtains the comoving Hubble radius of meVSL using Eqs.~\eqref{tkappaconstmpApp} and \eqref{H2meApp} 
\begin{align}
\frac{\tilde{c}}{a H} = \frac{\tilde{c}_0}{a H^{(\textrm{SM})}} \label{HR} \,.
\end{align}
As shown in the above equation~\eqref{HR}, the Hubble radius of meVSL is the same as that of SM.  To be a feasible model, we cannot propose that only the speed of light can vary, and in such a model, the Hubble radius may be the same as the SM.

\section{Direct probes of the dynamics of the expansion}
\label{sec:Dmeadotz} 

\subsection{Sandage Loeb effect}
\label{subsec:SLeff}

The time variation of the redshift of a source depends on the underlying cosmological model. The so-called Sandage-Loeb effect is the method that characterizes the change of the redshift for one source over time \citet{1962ApJ...136..319S,1998ApJ...499L.111L}.  Observationally, one needs to collect data from two light cones separated by a temporal period of $10-20$ years to find a change in temporal redshift $\Delta z_s / \Delta t_0$ as a function of the redshift of this source. Thus,  one only needs to use stable spectral lines and can reduce uncertainties from systematic or evolutionary effects.  To use the definition of the so-called RD, $\Delta z_s / \Delta t_0$,  one uses the observed source redshift changes during the time interval between the first and the second crest of light \citet{1998ApJ...499L.111L}
 \begin{align}
 \Delta z_s &\equiv z_s(t_0 + \Delta t_0) - z_s(t_0) = \frac{a(t_0 + \Delta t_0)}{a(t_s + \Delta t_s)} - \frac{a(t_0)}{a(t_s)} \nonumber \\ &\approx \Delta t_0  \left( \frac{\dot{a}(t_0)}{a(t_s)} -\frac{\dot{a}(t_s)}{a(t_s)} \frac{a(t_0)}{a(t_s)} \frac{\Delta t_s}{\Delta t_0} \right) \nonumber \\&\approx \Delta t_0 \left[ H_0 (1+z_s) - H(t_s) (1+z_s) (1+z_s)^{-1+b/4} \right] \nonumber \\
&= \Delta t_0 \left[ H_0 (1+z_s) - H(z_s)^{(\textrm{SM})} \right] \nonumber \\ 
&\equiv  \Delta t_0 H_0 \left[ (1+z_s) - E(z_s) \right] = \Delta z_s^{(\textrm{SM})} \label{DzLoeb} \,.
 \end{align}
The RD in the meVSL model is the same as that of SM. Thus, we cannot use the RD to probe the varying speed of light. This result is different from other VSL models \citet{2014PhLB..728...15B,2016PhRvD..93f3521S,2017PhRvD..95h4035S}. Both the CODEX (COsmic Dynamics and EXo-earth experiment) experiment \cite{2008MNRAS.386.1192L} proposed for the European-Extremely Large Telescope (EELT) and radio telescopes as Square Kilometer Array (SKA) \cite{2015aska.confE..27K} will grant access to direct measurements of the Hubble parameter up to redshifts $z \sim 5$. There has been increasing interest in SL observation due to its strong constraints power on the matter density parameter and its complementarity with other cosmological probes like SNe and BAO  \citet{2008PhRvD..77b1301U,2010PhLB..692..219J,2012PhR...521...95Q,2011JCAP...07..008G,2011PhRvD..84j4003M,2013ApJ...769..133N,2012PhRvD..86l3001M,2012ApJ...761L..26D,2015mgm..conf.1590M,2013JCAP...08..030A,2014MmSAI..85...13M,2013arXiv1311.6858Z,2015APh....62..195K,2015PhRvD..91f3006L,2015FrPhy..10j9501G,2016PhRvD..94d3001M,2016MNRAS.463L..61M,2017PhRvD..95j1301P,2018JCAP...04..002J,2018EPJC...78...11L,2018PhRvD..98b3517M,2019EPJC...79..267T,2019BAAS...51c..53E,2020JCAP...01..054J,2019arXiv190704495B,2019MNRAS.488.3607A,2020EPJC...80..304L,2019BAAS...51g.137E,2020MNRAS.492.2044C,2020JCAP...04..043L,2020NewA...8101425B,2021PhRvD.103b3537H,2021JCAP...12..042Q,2021PhRvD.103h1302H,2021PhRvL.126w1101K,2021arXiv210704868C,2021PhRvD.104l3527H,2021MNRAS.508L..53E,2021MNRAS.508.3620W,2022EJPh...43c5601M,2022LRR....25....6M,2022arXiv220305924C,2022MNRAS.513.5198C,2022PhRvD.106d3501K,2022MNRAS.514.5493D,2023NewA...9901940S,2022ApJ...940...16W,2022arXiv221013946L,2023MNRAS.518.2853R}.


\subsection{Cosmic chronometers}
\label{subsec:coschro}

The CC is to observe two passively evolving (elliptical) galaxies that were assumed to have formed at the same cosmic epoch but observed at different redshifts \citet{2002ApJ...573...37J}.  It is a model-independent method to measure the Hubble parameter as a function of redshift $H(z)$. The difference in their redshifts, $dz$, is obtained from spectroscopic surveys with high precision $\sigma_z \leq 0.001$.  Then the expansion rate (\textit{i.e.}, the Hubble parameter $H(z)$) is obtained from the differential age evolution of the Universe $dt$ in a given redshift interval ($dz$)
\begin{align}
H(z) \equiv \frac{\dot{a}}{a} &= - \frac{1}{1+z} \frac{dz}{dt} \approx - \frac{1}{1+z} \frac{\Delta z}{\Delta t} = H(z)^{(\textrm{SM})} (1+z)^{-b/4} \nonumber \\ &= H_0 E(z)^{(\textrm{SM})} (1+z)^{-b/4} \label{HzCC} \,,
\end{align}
where $E^{(\textrm{SM})}$ is given in Eq.~\eqref{H2meApp} and we assume that the differential redshift-time relation ($dz/dt \simeq \Delta z/\Delta t$) could be measured.  There are several methods to measure $\Delta t$. One can predict its age if the chemical composition of an ensemble of stars composing a simple stellar population is known. A different approach using a direct spectroscopic observable (the 4000 \AA \, break) known to be linearly related (at fixed metallicity) with the age of the stellar population has been used \cite{2011JCAP...03..045M}.  While many cosmological measurements rely on integrated distances, the CC determines the expansion rate $H(z)$ as a function of the redshift-time derivative $dz/dt$. Thus, this model-independent method has been used as a powerful discriminator to test different cosmological models \citet{2017ApJ...835..270W,2017MNRAS.467.3239R,2020ApJ...888...99W,2020ApJ...898...82M,2021ApJ...908...84V,2021MNRAS.506L...1D,2022ApJ...927..164B,2022ApJ...928L...4B,2022arXiv220902426B,2022EPJP..137.1341J,2022arXiv221203821A,2022arXiv221205731K,2022arXiv221214291L}. Thus, it is good to use for investigating VSL models \citet{2022JCAP...07..029R}. There are $31$ measurements of $H(z)$ by the CC between redshift $z = 0.07$ and $1.965$ \citet{2003ApJ...593..622J,2005PhRvD..71l3001S,2010JCAP...02..008S,2012JCAP...08..006M,2014RAA....14.1221Z,2015MNRAS.450L..16M,2016JCAP...05..014M,2022arXiv220505701J}. These are shown in Table \ref{tab:Hzall}. One can use these data to constrain the varying speed of light \cite{2022JCAP...07..029R}.

\begin{table} 
\begin{center}
\begin{tabular}{cccc}
\multicolumn{4}{c}{{}}\\
\hline \hline
$z$ & $H$ & $\sigma_{H}$ & ref\\
\hline
0.07 & 69.0 & 19.6 & \citet{2014RAA....14.1221Z} \\
0.09 & 72 & 12 &\citet{2003ApJ...593..622J}  \\
0.09 & 69 & 12 & \citet{2005PhRvD..71l3001S}\\
0.12 & 68.6 & 26.2 & \cite{2014RAA....14.1221Z} \\
0.17 & 83 & 8 & \citet{2005PhRvD..71l3001S}\\
0.179 & 75 & 4 & \citet{2012JCAP...08..006M}\\
0.199 & 75 & 5 & \citet{2012JCAP...08..006M}\\
0.20 & 72.9 & 29.6 & \citet{2014RAA....14.1221Z} \\
0.27 & 77 & 14 & \citet{2005PhRvD..71l3001S}\\
0.28 & 88.8 & 36.6 & \citet{2014RAA....14.1221Z} \\
0.352 & 83 & 14 & \citet{2012JCAP...08..006M}\\
0.3802 & 83 & 13.5 & \citet{2016JCAP...05..014M}\\
0.4 & 95 & 17 & \citet{2005PhRvD..71l3001S}\\
0.4004 & 77 & 10.2 & \citet{2016JCAP...05..014M}\\
0.4247 & 87.1 & 11.2 & \citet{2016JCAP...05..014M}\\
0.44497 & 92.8 & 12.9 & \citet{2016JCAP...05..014M} \\
\hline \hline
\end{tabular}
\begin{tabular}{cccc}
\multicolumn{4}{c}{{}}\\
\hline \hline
$z$ & $H$ & $\sigma_{H}$ & ref\\
\hline
0.4783 & 80.9 & 9 & \citet{2016JCAP...05..014M}\\
0.48 & 97 & 62 & \citet{2010JCAP...02..008S}\\
0.593 & 104 & 13 & \citet{2012JCAP...08..006M}\\
0.68 & 92 & 8 & \citet{2012JCAP...08..006M}\\
0.781 & 105 & 12 & \citet{2012JCAP...08..006M}\\
0.8 & 113.1 & 15.2 & \citet{2022arXiv220505701J}  \\
0.875 & 125 & 17 & \citet{2012JCAP...08..006M}\\
0.88 & 90 & 40 & \citet{2010JCAP...02..008S}\\
0.9 &  117 &  23 & \citet{2005PhRvD..71l3001S}\\
1.037 & 154 & 20 & \citet{2012JCAP...08..006M}\\
1.3 & 168 & 17 & \citet{2005PhRvD..71l3001S}\\
1.363 & 160 & 33.6 & \citet{2015MNRAS.450L..16M}\\
1.43 & 177 & 18 & \citet{2005PhRvD..71l3001S}\\
1.53 & 140 & 14 & \citet{2005PhRvD..71l3001S}\\
1.75 & 202 & 40 & \citet{2005PhRvD..71l3001S}\\
1.965 & 186.5 & 50.4 & \citet{2015MNRAS.450L..16M}\\
\hline \hline
\end{tabular}
\caption{$H(z)$ measurements (in units [km/s/Mpc]) used for the cosmological analysis, and their errors.}
\label{tab:Hzall}
\end{center}
\end{table}

We perform both the minimum $\chi^2$ analysis and the maximum likelihood one by using the current CC data in Tab.~\ref{tab:Hzall} to constrain the parameter $b$ of the meVSL model given in equation \eqref{HzCC}. 

\subsubsection{The minimum $\chi^2$}
\label{subsubsec:minchi}

For the $\chi^2$ analysis, we adopt other cosmological parameters ($H_0, \Omega_{m0}$) from the Planck mission  \citet{2020A&A...641A...6P} or Pantheon22 data  \citet{2022ApJ...938..110B} with $k = 0$ (\textit{i.e.}, a flat universe). We also analyze the data with the SM (\textit{i.e.}, $b =0$). The results are shown in Tab.~\ref{tab:chi2}.  When we adopt the Planck best-fit values for $(H_0, \Omega_{m0}) = (67.4 \text{km/s/Mpc}, 0.315)$, the SM ($b=0$) yields a good fit to the data ($\chi^2 = 14.97$ for $31$ degrees of freedom (dof)) with a goodness-of-fit (GoF) = $99\%$ (defined as GoF = $\Gamma(\nu/2\,,\chi^2/2)/\Gamma(\nu/2)$, where $\Gamma(\nu/2\,,\chi^2/2)$ is the upper  incomplete Gamma function, $\Gamma(\nu/2)$ is the Gamma function, and $\nu$ is the number of dof).  When we analyze the data for the meVSL model,  the best-fit value of $b$ is given by $-0.105 \pm 0.178$ within 1-$\sigma$ confidence level (CL) with $(\chi^2, \textrm{dof}) = (14.63,  30)$ to give GoF = $99\%$.  Thus, we can conclude that the current CC data is consistent with the SM if we adopt the Planck best-fit values for $(H_0, \Omega_{m0})$.  If the accuracies of foreseen CC observations are improved, then we might have a chance to distinguish the meVSL model from the SM. We repeat the same analysis by adopting Pantheon22 best-fit values for $(H_0, \Omega_{m0}) = (73.4 \text{km/s/Mpc}, 0.334)$.  In this case, the SM gives a much worse fit to the data with a GoF = $57 \%$.  However, the meVSL provides a good fit to the data (a GoF = $96 \%$) with a $1$-$\sigma$ CL values of the best-fit value $b = 0.584 \pm 0.184$.  Thus, we may conclude that the CC data prefer the meVSL to the SM if we adopt best-fit values for $(H_0, \Omega_{m0})$ from Pantheon22 data. We also repeat the same analysis by adopting only $\Omega_{m0} = 0.267$ from WMAP $7$-year data  \citet{2011ApJS..192...14J}.  We obtain almost the same GoF as other models.  In this model, we get a $1$-$\sigma$ CL $b = -0.318 \pm 0.329$ and $H_0 = 68.36 \pm 2.87$. This result is consistent with the WMPA $7$-year data $b=0$ and $H_0 = 71.0 \pm 2.5$. We repeat the same analysis for a chosen value of $\Omega_{m0} = 0.340$. The results of this model are pretty similar to those of others.  The GoF is $98\%$ with the $1$-$\sigma$ CL for $b$ and $H_0$ are $0.108 \pm 0.331$ and $68.07 \pm 2.86$, respectively.  Thus, we might conclude that the current CC data show $H_0 \approx 68$ without depending on the value of $\Omega_{m0}$. However,  the current CC data cannot constrain the value of $\Omega_{m0}$ when we fix the value of $H_0$.  

\begin{table*}
\centering
\caption{A simple $\chi^2$ analysis for the meVSL model using CC data by adapting $(H_0\,,\Omega_{m 0})$ values from Planck18 \citet{2020A&A...641A...6P} and Pantheon22 \citet{2022ApJ...938..110B} , respectively.  The best-fit values for the $b$ exponent are $b = -0.105 \pm 0.178$ and $0.584 \pm 0.184$, respectively.  Also, we show results for adopting $\Omega_{m 0} = 0.267$ from WMAP 7 year data \citet{2011ApJS..192...14J} and for $\Omega_{m 0} = 0.340$.  }
\label{tab:chi2}
\begin{tabular}{ |c|c|c|c|c|c|} 
 \hline
($H_0\,,\Omega_{m0}$) & $\chi^2$/dof & $\bar{b}$ & $1$-$\sigma$ & GoF (\%) & ref \\ 
\hline
\multirow{2}{*}{($67.4\,,0.315$)} & $14.97/31$ & $0$ &  & $99$ & \multirow{2}{*}{\citet{2020A&A...641A...6P} }\\ 
& $14.63/30$ & $-0.105$ & $0.178$ & $99$ & \\  
\multirow{2}{*}{($73.4\,,0.334$)} & $29.04/31$ & $0$  & $$ & $57$ &  \multirow{2}{*}{\citet{2022ApJ...938..110B}} \\ 
& $17.82/30$ & $0.584$  & $0.184$ & $96$ & \\ 
 \hline
$\Omega_{m0}$ & $\chi^2$/dof & $b$ & $H_0$ & GoF (\%) & ref \\ 
\hline
$0.267$ & $14.60/29$ & $-0.318 \pm 0.329$ & $68.36 \pm 2.87$ & $99$ & \citet{2011ApJS..192...14J} \\  
$0.340$ & $14.54/29$ & $0.108 \pm 0.331$ & $68.07 \pm 2.86$ & $99$ & \\  
\hline
\end{tabular}
\end{table*}

\begin{figure*}
\centering
\vspace{1cm}
\begin{tabular}{cc}
\includegraphics[width=0.49\linewidth]{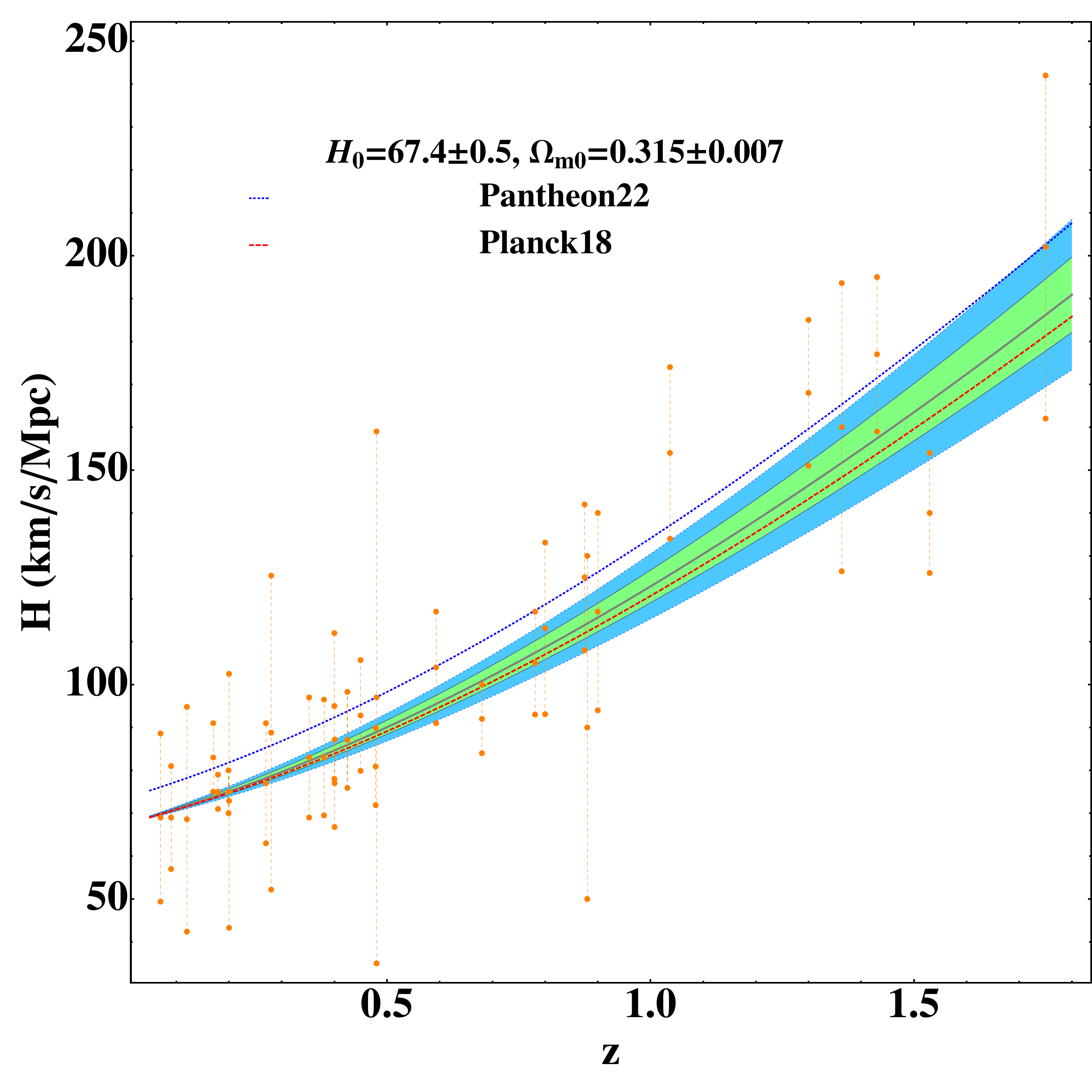} &
\includegraphics[width=0.49\linewidth]{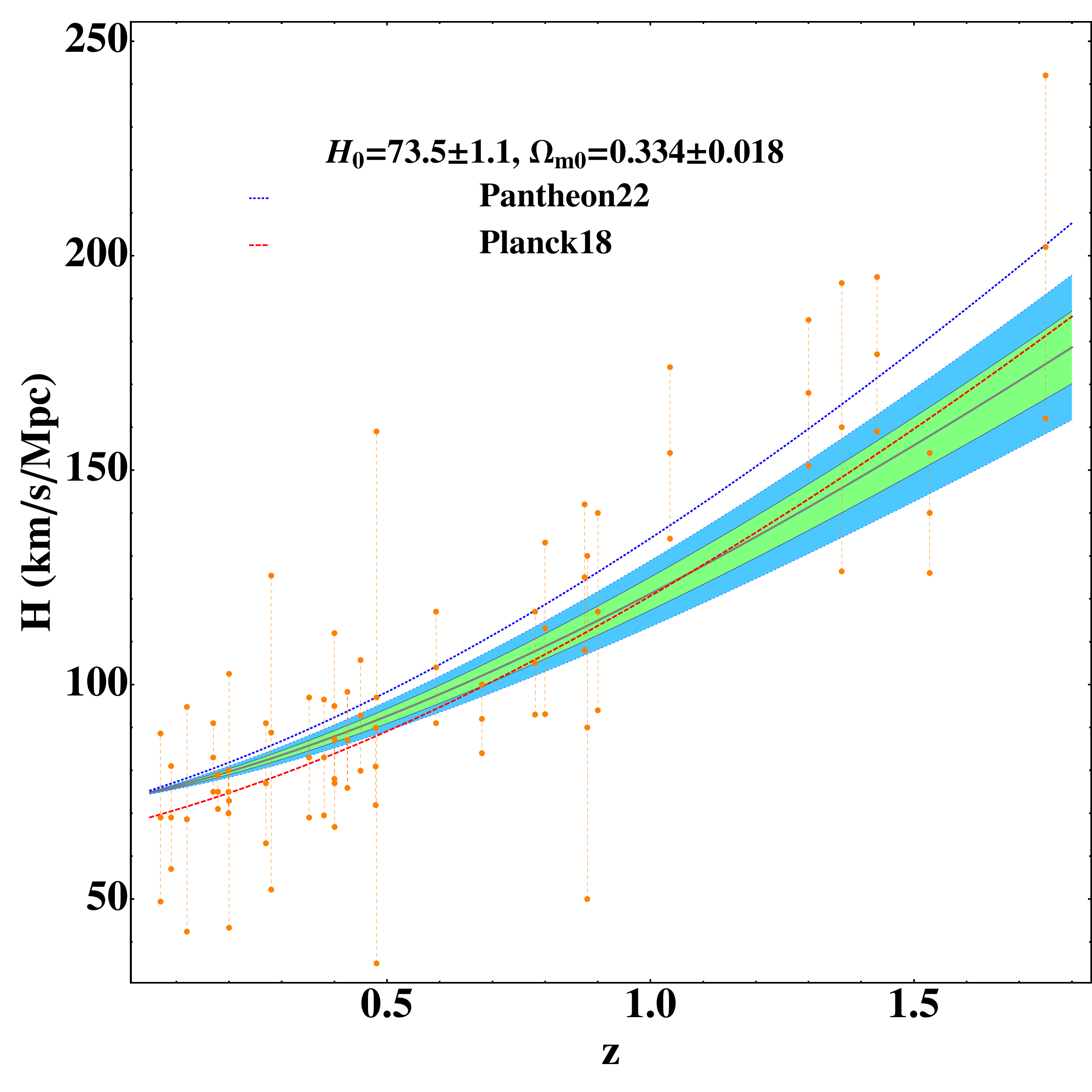}
\end{tabular}
\vspace{-0.5cm}
\caption{Results obtained from the minimum $chi^2$ analysis. Hubble parameters of the SM versus those of the meVSL model given in \eqref{HzCC} with the $31$ CC data.  a) The dashed (dotted) line denotes the SM $H(z)$ when we adopt best-fit values of ($H_0,\Omega_{m0}$) from the Planck mission (Pantheon22).  The best-fit meVSL model with $b = -0.105 \pm 0.178$ (adopting the Planck best-fit ($H_0,\Omega_{m0}$) values) is depicted as a solid line and bright (dark) shaded area. within the $1 \,(2)$-$\sigma$ CL.  b) The best-fit meVSL model with $b = 0.584 \pm 0.184$ (adopting the Pantheon22 best-fit ($H_0,\Omega_{m0}$) values) is depicted as a solid line and bright (dark) shaded area. within the $1 \,(2)$-$\sigma$ CL. } \label{fig-1}
\vspace{1cm}
\end{figure*}

We also show these results in Fig.~\ref{fig-1}. In the left panel of Fig.~\ref{fig-1}, we depict the $31$ CC data with their error bars indicated by vertical lines.  The dashed and dotted lines show the evolutions of SM Hubble parameters when we adopt best-fit values of ($H_0,\Omega_{m0}$) from the Planck mission and Pantheon22 data, respectively.  The solid line and the bright (dark) shaded regions represent both the best fit and the $1$ ($2$)-$\sigma$  constraints from the CC data with adopting ($H_0,\Omega_{m0}$) from the Planck mission. As shown in the figure,  the $1$-$\sigma$ results are well matched with those of the SM (the dashed line).  We also depict the $1$ ($2$)-$\sigma$ constraints from the CC data by adopting ($H_0,\Omega_{m0}$) from the Pantheon22 data in the right panel of Fig.~\ref{fig-1}. The evolution of the SM Hubble parameter with the Pantheon22 best-fit values of ($H_0,\Omega_{m0}$) is denoted by the dotted line.  It shows the deviation from the $2$-$\sigma$ constraints represented by the dark-shaded regions.

\subsubsection{The maximum likelihood}
\label{subsubsec:maxlike}

We perform the maximum likelihood analysis in this subsection by marginalizing over the informative prior for $(H_0\,,\Omega_{m 0})$ instead of fixing those values. Because they might not represent their best fits to those CC data. Thus, we assume that the prior distribution of $H_0$ ($\Omega_{m0}$) is Gaussian with the mean $\bar{H}_0$ ($\bar{\Omega}_{m0}$) and the standard deviation $\sigma_{H_0}$ ($\sigma_{\Omega_{m0}}$)
\begin{align} &P(X_0) = \frac{1}{\sqrt{2 \pi \sigma^2_{X_0}}} e^{-(X_0 - \bar{X}_0)^2/(2 \sigma^2_{X_0})} \,,  \nonumber \\ 
&\textrm{where} \, X_0 = H_0 \,\, \textrm{or} \,\, \Omega_{m0} \label{PX0} \,. \end{align}
($\bar{H}_0$\,, $\sigma_{H_0}$) = ($67.4$\,,$0.5$) and ($\bar{\Omega}_{m0}$\,,$\sigma_{\Omega_{m0}}$) = ($0.315$\,,$0.007$) for Planck mission and ($\bar{H}_0$\,, $\sigma_{H_0}$) = ($73.5$\,,$1.1$) and ($\bar{\Omega}_{m0}$\,,$\sigma_{\Omega_{m0}}$) = ($0.334$\,,$0.018$) for Pantheon22. Then, we can build the posterior likelihood function $\mathcal{L}_{H} (b)$ by marginalizing over both $H_0$ and $\Omega_{m0}$ 
\begin{align}
\mathcal{L}_{H}(b) = \int_{0}^{\infty} e^{-\chi^2(H_0\,,\Omega_{m0}\,,b)} P(H_0) dH_0 P(\Omega_{m0}) d \Omega_{m0} \label{calL} \,,
\end{align}
We maximize the likelihood with respect to the parameter $b$ to obtain the best-fit parameter value of it.  We also evaluate the maximum likelihood values of the SM both for the Planck mission and for the Patheon22 data to compare model preferences over data. We obtain the Akaike information criterion (AIC) and the Bayesian information criterion (BIC) for the meVSL model and SM. 
\begin{align} 
\textrm{AIC} &= -2 \ln \mathcal{L} + 2 k \,, \nonumber \\ 
\textrm{BIC} &= -2 \ln \mathcal{L} + k \ln N \label{AICBIC} \,, \end{align}
where $\mathcal{L}$ is the maximum likelihood, $k$ is the number of parameters of the model, and $N$ is the number of datapoints used in the analysis. 

We analyze the data for both the SM and the meVSL. The results are summarized in Tab.~\ref{tab:maxli}.  When we marginalize over the prior for the Planck mission, the SM ($b=0$) yields the value of AIC $15.03$ and the same BIC because there is no free parameter.  When we analyze the data for the meVSL model,  the best-fit value of $b$ is given by $-0.102 \pm 0.329$ within 1-$\sigma$ confidence level (CL) with AIC $= 16.73$ and BIC $= 18.16$.  Thus, we can conclude that the current CC data prefer the SM if we adopt priors on cosmological parameters from the Planck mission.  We repeat the same analysis by adopting prior from Pantheon22 data.  In this case, the AIC (BIC) of the SM is $24.90$.  The meVSL provides AIC = $19.91$ and BIC = $21.35$ with a $1$-$\sigma$ CL values of the best-fit value $b = 0.532 \pm 0.343$.  Thus, we may conclude that the CC data prefer the meVSL to the SM if we marginalize with Pantheon22 data. 

\begin{table*}
\centering
\caption{The maximum likelihood analysis for the meVSL model and SM using CC data by adapting Gaussian prior on cosmological parameters from Planck18 \citet{2020A&A...641A...6P} and Pantheon22 \citet{2022ApJ...938..110B} , respectively.  The best-fit values for the $b$ exponent are $b = -0.102 \pm 0.329$ and $0.532 \pm 0.343$, respectively. }
\label{tab:maxli}
\begin{tabular}{ |c|c|c|c|c|c|c|} 
 \hline
($\bar{H}_0$\,, $\bar{\Omega}_{m0}$) & ($\sigma_{H_0}$\,,$\sigma_{\Omega_{m0}}$)& $\bar{b}$ & $1$-$\sigma$ & AIC & BIC & ref \\ 
\hline
\multirow{2}{*}{($67.4\,,0.315$)} & \multirow{2}{*}{($0.5\,,0.007$)} & $0$ &  & $15.03$ & $15.03$ & \multirow{2}{*}{\citet{2020A&A...641A...6P} }\\ 
&  & $-0.102$ & $0.329$ & $16.73$ & $18.16$ & \\  
\multirow{2}{*}{($73.4\,,0.334$)} & \multirow{2}{*}{($1.1\,,0.018$)} & $0$  & $$ & $24.90$ & $24.90$&  \multirow{2}{*}{\citet{2022ApJ...938..110B}} \\ 
&  & $0.532$  & $0.343$ & $19.91$ & $21.35$ & \\ 
 \hline
\end{tabular}
\end{table*}

\begin{figure*}
\centering
\vspace{1cm}
\begin{tabular}{cc}
\includegraphics[width=0.49\linewidth]{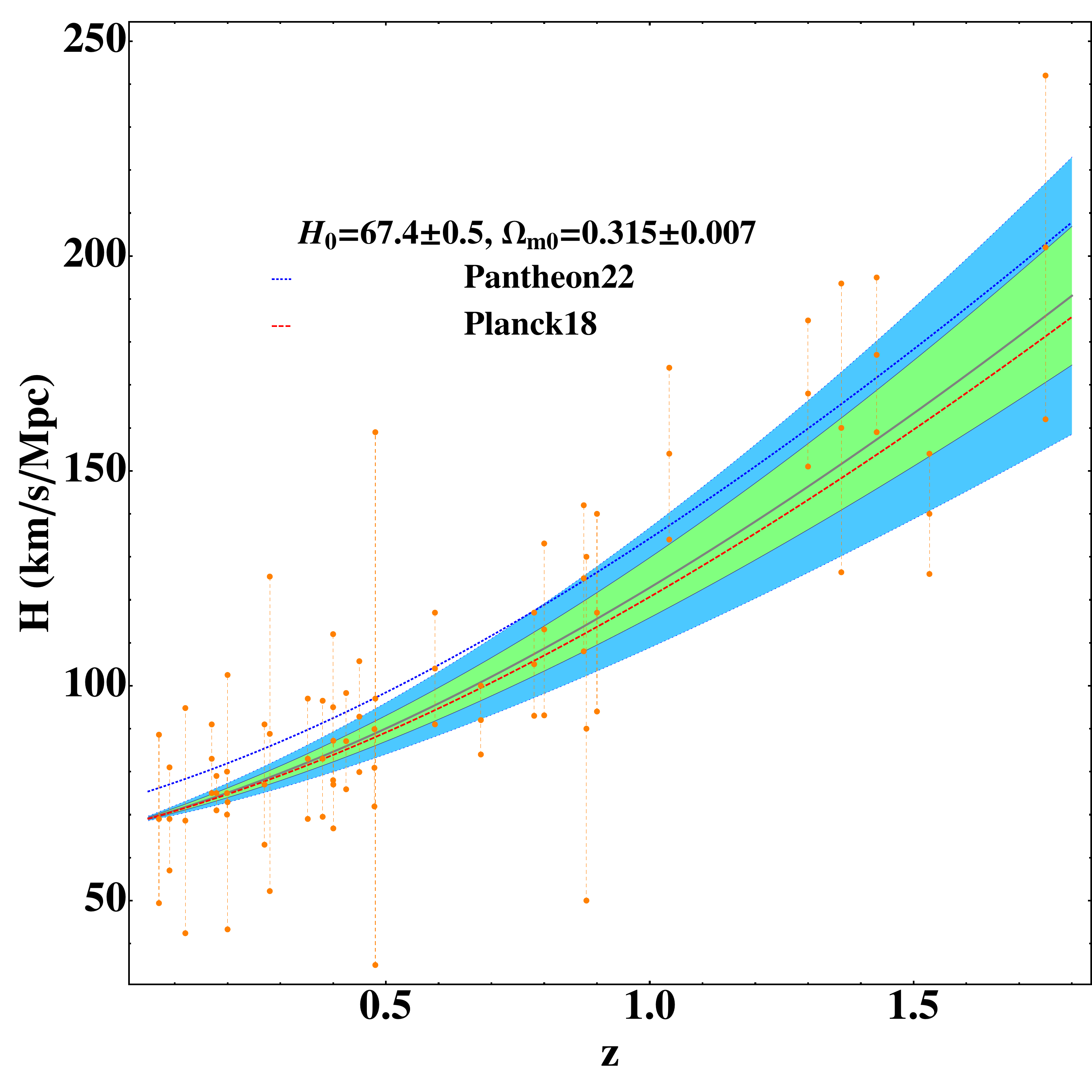} &
\includegraphics[width=0.49\linewidth]{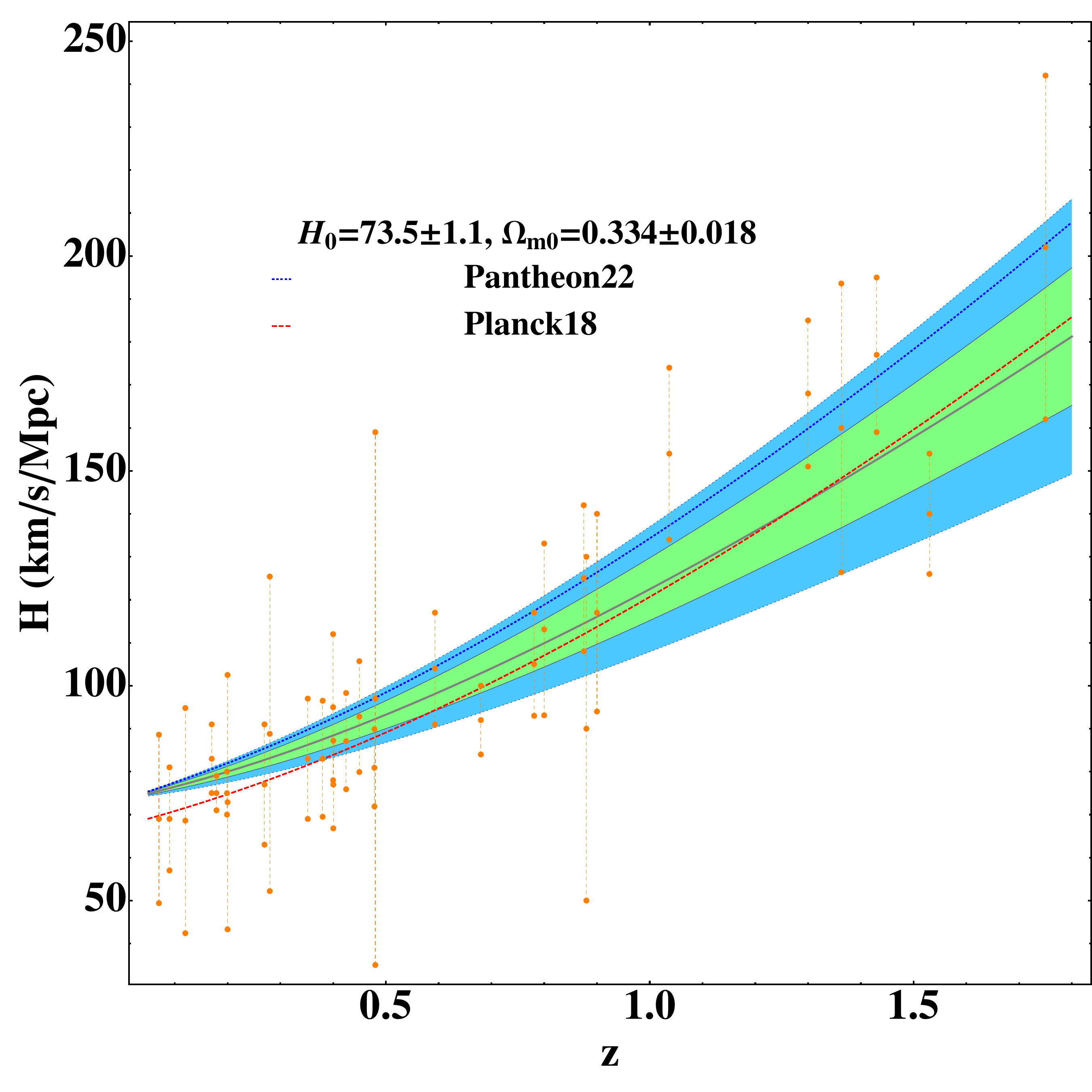}
\end{tabular}
\vspace{-0.5cm}
\caption{Results obtained from the maximum likelihood analysis.  a) The dashed (dotted) line denotes the SM $H(z)$ when we adopt best-fit values of ($H_0,\Omega_{m0}$) from the Planck mission (Pantheon22).  The best-fit meVSL model with $b = -0.102 \pm 0.329$ (adopting the Planck prior) is depicted as a solid line and bright (dark) shaded area within the $1 \,(2)$-$\sigma$ CL.  b) The best-fit meVSL model with $b = 0.532 \pm 0.343$ (adopting the Pantheon22 prior) is depicted as a solid line and bright (dark) shaded area. within the $1 \,(2)$-$\sigma$ CL..} \label{fig-2}
\vspace{1cm}
\end{figure*}

We also show these results in Fig.~\ref{fig-2}. In the left panel of Fig.~\ref{fig-2},  the dashed and dotted lines show the evolutions of SM Hubble parameters when we adopt best-fit values of ($H_0,\Omega_{m0}$) from the Planck mission and Pantheon22 data, respectively.  The solid line and the bright (dark) shaded regions represent both the best fit and the $1$ ($2$)-$\sigma$  constraints from the CC data with marginalization over Gaussian prior on cosmological parameters ($H_0,\Omega_{m0}$) from the Planck mission. As shown in the figure,  the $2$-$\sigma$ results are well matched with the SM (the dashed line) except for the deviation from the Pantheon22 prediction at low $z$.  We also depict the $1$ ($2$)-$\sigma$ constraints from the CC data by adopting ($H_0,\Omega_{m0}$) from the Pantheon22 data in the right panel of Fig.~\ref{fig-2}. The evolution of the SM Hubble parameter with the Planck best-fit values of ($H_0,\Omega_{m0}$) deviates from 2-$\sigma$ at low $z$.  

\section{Conclusions}
\label{sec:Conc}

We have driven a formula for cosmic chronometers in the context of the minimally extended varying speed of light theory. We use the current cosmic chronometer data to constrain it.  If we adopt the values of the current Hubble parameter $H_0$ and the matter density contrast $\Omega_{m0}$ in the Planck mission,  we cannot distinguish between the meVSL model and the standard one. If the errors in the foreseen data decrease,  there may be an opportunity to discern between the model.  Interestingly, the current cosmic chronometer data already shows the deviation of the meVSL model from the standard model if the values ($\Omega_{m0}, H_0$) are adopted from the Pantheon data. The upcoming cosmic chronometer data from Euclid,  DESI,  J-PAS, and SKA will probably be used as an independent test of the meVSL model because it might show a difference from the standard model. 

\section*{Acknowledgements}

SL is supported by Basic Science Research Programs through the National Research Foundation of Korea (NRF) funded by the Ministry of Science, ICT, and Future Planning (Grant No. NRF-2017R1A2B4011168 and No. NRF-2019R1A6A1A10073079). 

\section*{Data Availability}

 The authors confirm that the data supporting the findings of this study are available within the references and their supplementary materials. No new data were created or analyzed in this study.



\bibliographystyle{mnras}
\bibliography{example} 








\bsp	
\label{lastpage}
\end{document}